\documentclass[amsmath,amssymb,amsfonts,showkeys]{revtex4}
 
\usepackage{balance}
\usepackage[normalem]{ulem}
 \usepackage{algorithmic}
\usepackage{graphicx}
\usepackage{textcomp}
\usepackage{xcolor}
\usepackage{microtype}
 
\usepackage{braket}
\usepackage{soul}
\usepackage{booktabs}
\usepackage[bookmarks=true,breaklinks=true,letterpaper=true,colorlinks,linkcolor=black,citecolor=blue,urlcolor=black]{hyperref}
\usepackage[capitalise]{cleveref}

\usepackage{adjustbox}

\usepackage{bm}
\usepackage{url}
\usepackage[linewidth=1pt]{mdframed}

\usepackage{tikz}

\usepackage{mathtools}

\usepackage{caption}
\usepackage{subcaption}
 \usepackage{tabularx}

\begin{document}

\title{Demonstration of a CAFQA-bootstrapped Variational Quantum Eigensolver on a trapped-ion quantum computer}
 
\author{Qingfeng Wang}

\email{q.kee.wang@gmail.com}
 
\affiliation{Duke Quantum Center, Departments of Electrical and Computer Engineering and Physics, Duke University, Durham, USA\\
 Chemical Physics Program and Institute for Physical Science and Technology, University of Maryland, College Park, Maryland, USA}

\author{Liudmila Zhukas}
 \affiliation{%
  Duke Quantum Center, Departments of Electrical and Computer Engineering and Physics, Duke University, Durham, USA}

\author{Qiang Miao}
 \affiliation{ Duke Quantum Center, Duke University,Durham,USA}

\author{Aniket S. Dalvi}
 \affiliation{ Department of Electrical and Computer Engineering, Duke University,Durham,USA }

\author{Peter J. Love}
 \affiliation{ Department of Physics and Astronomy, Tufts University,Medford,USA}

\author{Christopher Monroe}
 \affiliation{ Duke Quantum Center, Departments of Electrical and Computer Engineering and Physics, Duke University,Durham,USA}

\author{Frederic T. Chong}
 \affiliation{ Department of Computer Science, University of Chicago,Chicago,USA}

\author{Gokul Subramanian Ravi}
 \affiliation{ Department of Computer Science and Engineering, University of Michigan,Ann Arbor,USA}

\begin{abstract}
To enhance the variational quantum eigensolver (VQE), the CAFQA method can utilize classical computational capabilities to identify a better initial state than the Hartree-Fock method. Previous research has demonstrated that the initial state provided by CAFQA recovers more correlation energy than that of the Hartree-Fock method and results in faster convergence.
In the present study, we advance the investigation of CAFQA by demonstrating its advantages on a high-fidelity trapped-ion quantum computer located at the Duke Quantum Center---this is the first experimental demonstration of CAFQA-bootstrapped VQE on a TI device and on any academic quantum device.
In our VQE experiment, we use LiH and BeH$_2$ as test cases to show that CAFQA achieves faster convergence and obtains lower energy values within the specified computational budget limits.
To ensure the seamless execution of VQE on this academic device, we develop a novel hardware-software interface framework that supports independent software environments for both the circuit and hardware end. This mechanism facilitates the automation of VQE-type job executions as well as mitigates the impact of random hardware interruptions.
This framework is versatile and can be applied to a variety of academic quantum devices beyond the trapped-ion quantum computer platform, with support for integration with customized packages.
\end{abstract}
 

\maketitle

\section{Introduction}

\noindent\textbf{\emph{Quantum computing in the pre-fault-tolerant era:}} Quantum computing is anticipated to address numerous classically intractable problems using proven efficient algorithms, such as Shor's factoring algorithm ~\cite{Shor_1997} and Grover's search algorithm ~\cite{Grover96afast}. However, the successful execution of these algorithms necessitates a fault-tolerant quantum computer, which the community generally believes is still many years away ~\cite{preskill2018quantum}.
Consequently, significant attention has been directed towards algorithms capable of performing practically useful quantum applications in the pre-fault-tolerant era \cite{kim2023evidence}. Among the most promising of these algorithms is the variational quantum algorithm (VQA), a type of classical-quantum hybrid heuristic algorithm that can leverage gate-based noisy quantum computers ~\cite{peruzzo2014variational, farhi2014quantum,moll2018quantum}. Since the inception of this algorithm, substantial efforts have been made to enhance the execution capability of VQAs on near-term noisy quantum devices. This includes methods to reduce measurement overhead, including optimal schemes for measurement grouping~\cite{gokhale2019minimizing,verteletskyi2020measurement,huggins2022nearly,bonet2020nearly,choi2023measurement,choi2023fluid,yen2021cartan,choi2022improving}, reduction by grouping into anticommuting sets\cite{izmaylov2019unitary,zhao2020measurement} and by shadow tomography techniques~\cite{AaronsonSTOC2020,HuangNatPhys2020,ChenPRXQ2021,zhao2021fermionic,EnshanQ2022}. Another area of improvement of VQE is methods to improve the ansatz \cite{wang2023ever,wang2021resource,adaptvqe} and to mitigate errors \cite{czarnik2020error,Rosenberg2021,barron2020measurement,botelho2021error,wang2021error,takagi2021fundamental,temme2017error,li2017efficient,giurgica2020digital,ding2020systematic,smith2021error,ravi2021vaqem,ravi2023navigatingdynamicnoiselandscape,dangwal2024varsawapplicationtailoredmeasurementerror,zhang2024disqdynamiciterationskipping}. For recent reviews see~\cite{cerezo2021variational, cai2023quantum, ElbenRev2022}.

\begin{figure}
     \centering
    \begin{minipage}{1\textwidth}
        \centering
         \subfloat[VAQ space break down]{
         \includegraphics[height=0.32\textwidth]{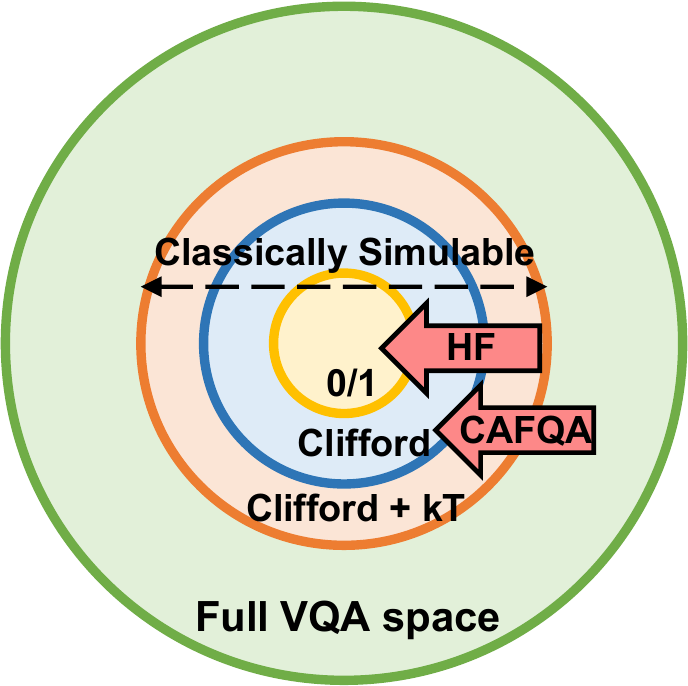}}
          \subfloat[CAFQA-compatible ansatz]{\includegraphics[height=0.32\textwidth]{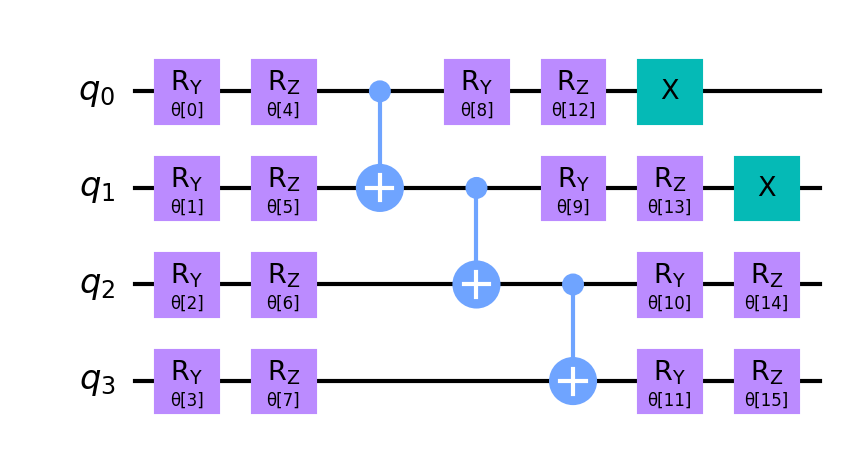}    }
            \caption{
        (a) (Imported from our prior work Fig. 2 \cite{ravi2022cafqa}.) 
        The plot demonstrates that the parameter space can be explored using different methods. 
        Here 0/1 represents the binary product state and T denotes the quantum $T$ gate.
        This concept is further illustrated by a CAFQA-compatible ansatz circuit in (b). 
        The circuit in (b) becomes Clifford and can be explored via CAFQA if variables are selected from a discrete set, $\theta[i]\pmod{ 2\pi}=\{0,\pi/2, \pi, 3\pi/2 \}$. 
        The $X$ gates are added explicitly to indicate that the circuit is reduced to generate a Hartree-Fock initial state with the first two orbitals occupied by electrons when $\theta[i] = 0$ for all $i$.  
        The parameter space become a classical intractable VQA space when variables are allowed to be arbitrary real numbers.
        }
        \label{fig:space}
        \label{fig:double}
        \label{fig:hardware-efficient}
    \end{minipage}
\end{figure}

\noindent\textbf{\emph{VQA assissted by CAFQA:}}
In previous work, CAFQA (\underline{C}lifford \underline{A}nsatz \underline{F}or \underline{Q}uantum \underline{A}ccuracy) was proposed as an effective method for initializing the VQA ansatz with the aid of classical support~\cite{ravi2022cafqa, bhattacharyya2024optimalcliffordinitialstates}. This method leverages classical computing capabilities, as Clifford circuits can be efficiently simulated on a classical computer in polynomial time~\cite{gottesman1998heisenberg,aaronson2004improved}. The study demonstrated that CAFQA can often identify an initialization configuration that is equal to or superior to the traditional Hartree-Fock (HF) initialization, as it encompasses a broader classically simulable region compared to HF initialization.
\cref{fig:space} (a) provides a comparison of the covered parameter spaces for both approach, and more details can be found in our prior work~\cite{ravi2022cafqa}. A brief explanation of VQA within the CAFQA formalism is provided in \cref{sec:Real-VQA}, and the CAFQA framework is summarized in \cref{sec:cafqa-framework}.

\noindent\textbf{\emph{CAFQA demonstration challenge:}}
Testing the CAFQA method necessitates the use of a low-noise quantum device. If the quantum device exhibits excessive noise, the VQA may fail to perform effectively, even with optimal initial states. For instance, significant noise can distort the system to such an extent that a state optimal in a noise-free scenario becomes a noisy local minimum on the device, which can be difficult to escape. Therefore, it is essential to identify and utilize machines with low noise levels for testing these methods. Among various platforms, such as superconducting and neutral atom quantum computers, trapped ion (TI) devices are particularly suitable for testing VQA circuits due to their low error rates and full connectivity.
We provide a brief background on running VQA with trapped-ion quantum computers in \cref{sec:background-TI} and offer more technical details about the specific TI device used in \cref{sec:trapped-ion}.
Additionally, testing on an academic device is crucial as it facilitates the development of a comprehensive hardware-software interface to support this research. In addition to classical support methods such as CAFQA, the successful execution of VQA on actual quantum devices requires full stack support, which is the primary focus of this work. We introduce the background for the hardware-software interface in \cref{sec:bg-api} and discuss the specific problems encountered and their implementation in \cref{sec:api}.

\vspace{0.2in}
\noindent\textbf{\emph{Key contributions:}}
Enabled by our novel hardware-software interfacing framework, we demonstrate CAFQA-bootstrapped VQE on our trapped-ion quantum computer for the LiH and BeH$_2$ molecules. 
The methodology is detailed in \cref{sec:CAFQA-exp} and additional discussion is provided  in \cref{sec:discussion}. We summarize the major contribution as follows:

\begin{enumerate}
    \item \textbf{\emph{API framework:}}
We have developed a comprehensive hardware-software stack to support VQA on our trapped-ion (TI) device, integrating state-of-the-art classical initialization support through CAFQA. The API facilitates streamlined and efficient VQA experiments, even in the presence of random hardware interruptions. This framework is adaptable to other quantum platforms as well and offers the flexibility for integrating other customized packages across the entire stack.
\item \textbf{\emph{CAFQA experiment result:}}
Assisted by high-fidelity gates on our trapped-ion quantum computer, we demonstrated that the advantage of CAFQA initialization persists in experimental validation. 
Through two test cases, we observed faster and more accurate VQE convergence when starting from CAFQA initialization compared to HF initialization, and absolute energy estimates closer to the ideal ground state energy. 
To our knowledge, this is the first experimental demonstration of CAFQA-bootstrapped VQE on a TI device and on any academic quantum device.
\end{enumerate}

\begin{figure*}[t]
\centering

\includegraphics[width=\textwidth,trim={0cm 0cm 0cm 0cm}]{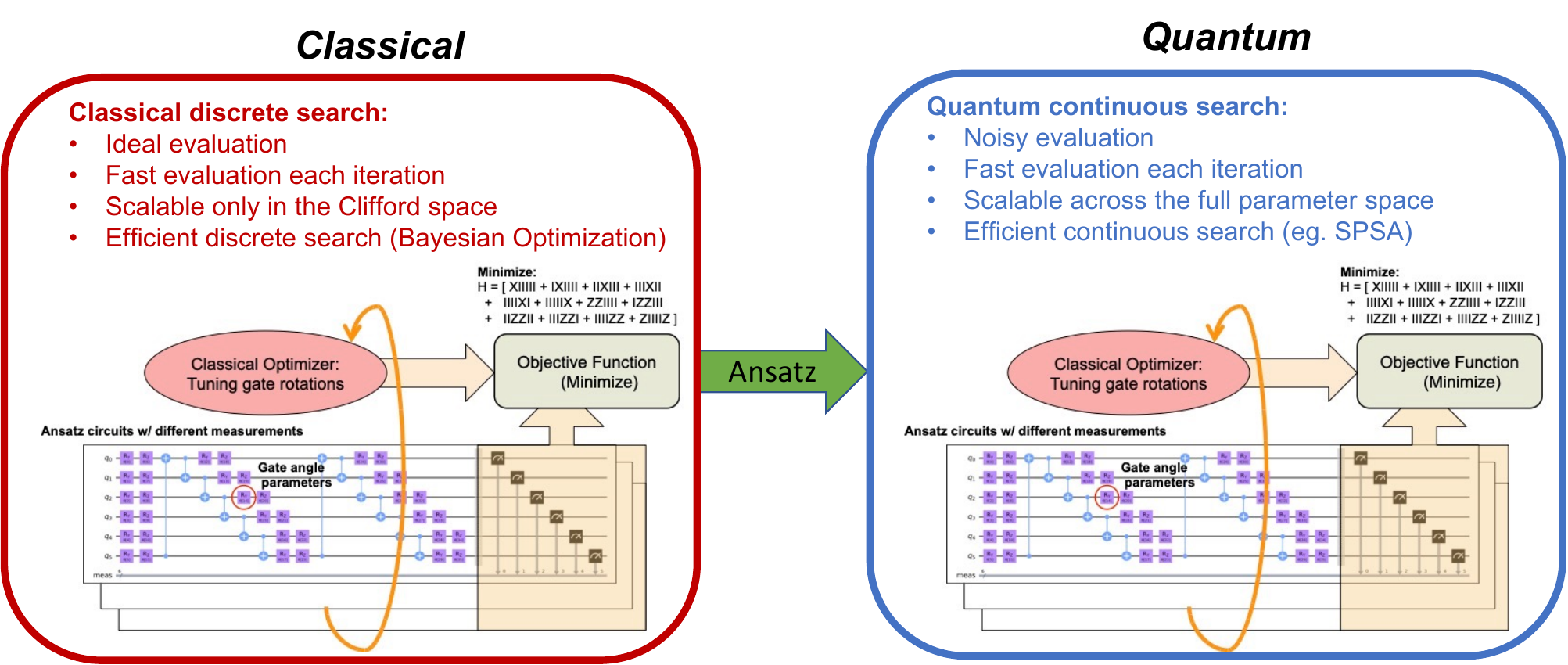}

\caption{(Imported from our prior work  \cite{ravi2022cafqa}.) 
The left red box represents the module for CAFQA discrete search on a classical computer using a Clifford circuit. The right blue box represents the traditional VQA optimization with initialization parameters derived from the CAFQA search outcome.}
\label{fig:qafca-intro}
\end{figure*}

 \section{Background and Motivation}

\subsection{VQA under CAFQA formalism}
 \label{sec:Real-VQA}
 Here we briefly summarize the formalism for VQA. 
For a more detailed description, refer to our prior work \cite{ravi2022cafqa} and a recentr review~\cite{cerezo2021variational}.
VQA can generally be formulated to minimize a cost function  of the form 
\begin{align*}
f_i=\langle \psi_0 | U^\dagger(\vec{\theta}_i) H U(\vec{\theta}_i) |\psi_0\rangle, 
\end{align*}
at step $i$,
where $U(\vec{\theta}_i) $ is the  \textbf{parametrized ansatz} evolved on a quantum computer with parameter vector $\vec{\theta}_i$ at step $i$.
Here $|\psi_0\rangle$ denotes the initial state of the quantum computer, typically initialized as the zero state.
The \textbf{initial state of the quantum circuit}, $U(\vec{\theta}_0)|\psi_0\rangle$, is formed using the parametrized ansatz $U(\vec{\theta}_0)$.
For example, in a CAFQA-compatible ansatz shown in  \cref{fig:hardware-efficient} (b),
$U(\vec{\theta}_0)|\psi_0\rangle$ 
represents the Hartree-Fock initialization with $\vec{\theta}_i = \bm{0}$.
Meanwhile,  it represents the CAFQA initialization if
$\vec{\theta}_i[j]\pmod{2\pi} \in \{0,\pi/2, \pi, 3\pi/2 \}$.
The Hamiltonian $H$ encodes the information of the system.
Here we consider a Hamiltonian consists of a summation of weighted Pauli strings.
In the context where $H$ represents the electronic interaction of a molecular system, and the cost function is the energy of the molecule, VQA is often referred to as VQE (Variational Quantum Eigensolver) \cite{peruzzo2014variational}.

\subsection{CAFQA framework}
\label{sec:cafqa-framework}

\noindent {\bf \emph{CAFQA procedure overview:}}
Here we summarize the proposal of the CAFQA method proposed in our previous work \cite{ravi2022cafqa}.
CAFQA performs parameter search using a classical computer over a discrete set of variables. This search is conducted using a CAFQA-compatible ansatz, such as the one illustrated in \cref{fig:hardware-efficient} (b), ensuring that the circuit remains Clifford and therefore classically simulable \cite{gottesman1998heisenberg}.
Upon completion of the CAFQA search, the optimal variables identified are transferred to VQA as the starting point for the subsequent continuous-variable optimization process. This procedure is depicted in \cref{fig:qafca-intro}.

\begin{figure}
\centering

 \centering
 
 \subfloat[LiH Energy]{\includegraphics[height=0.29\textwidth]{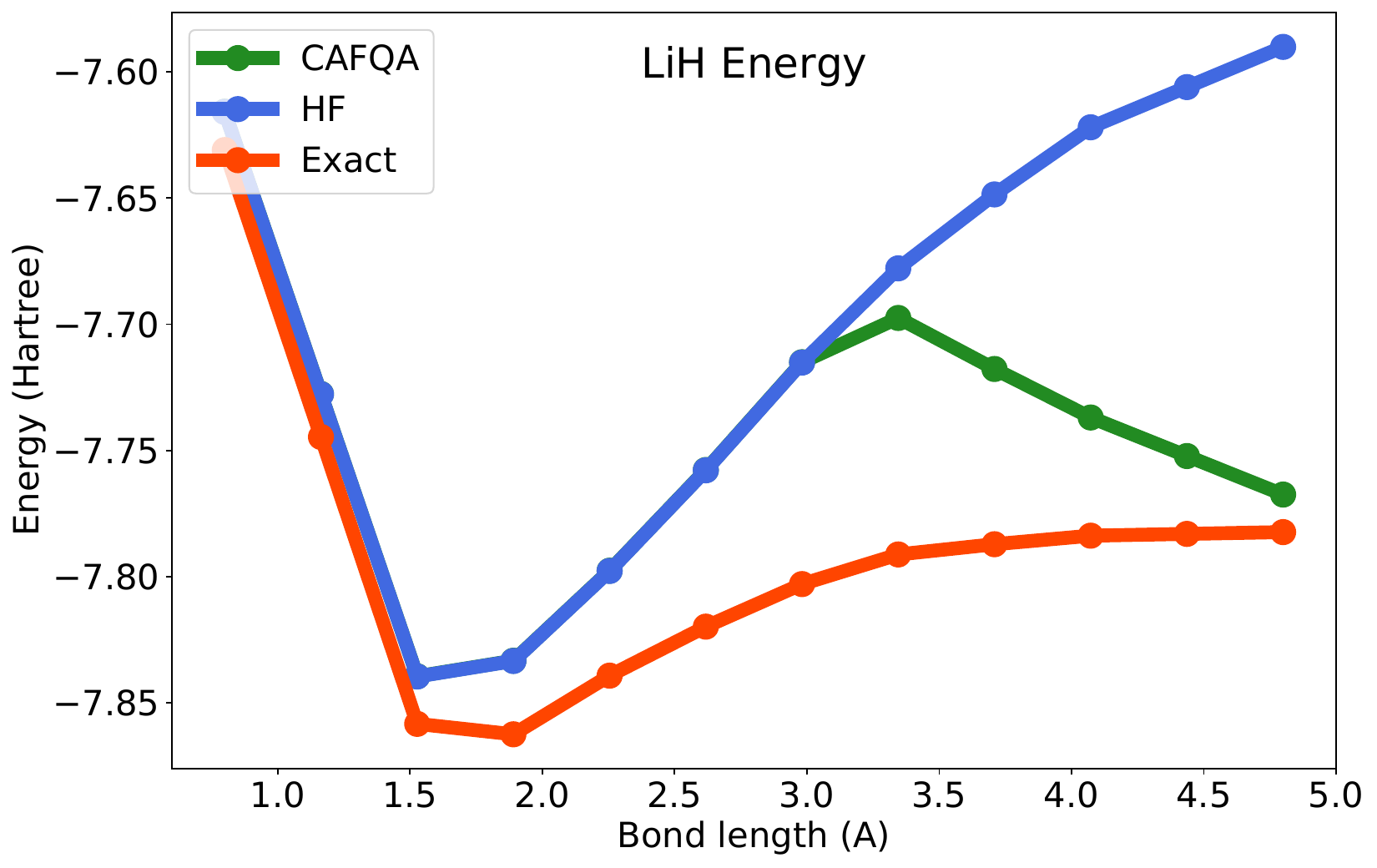}}
 
\subfloat[LiH VQA evolution]{\includegraphics[height=0.29\textwidth]{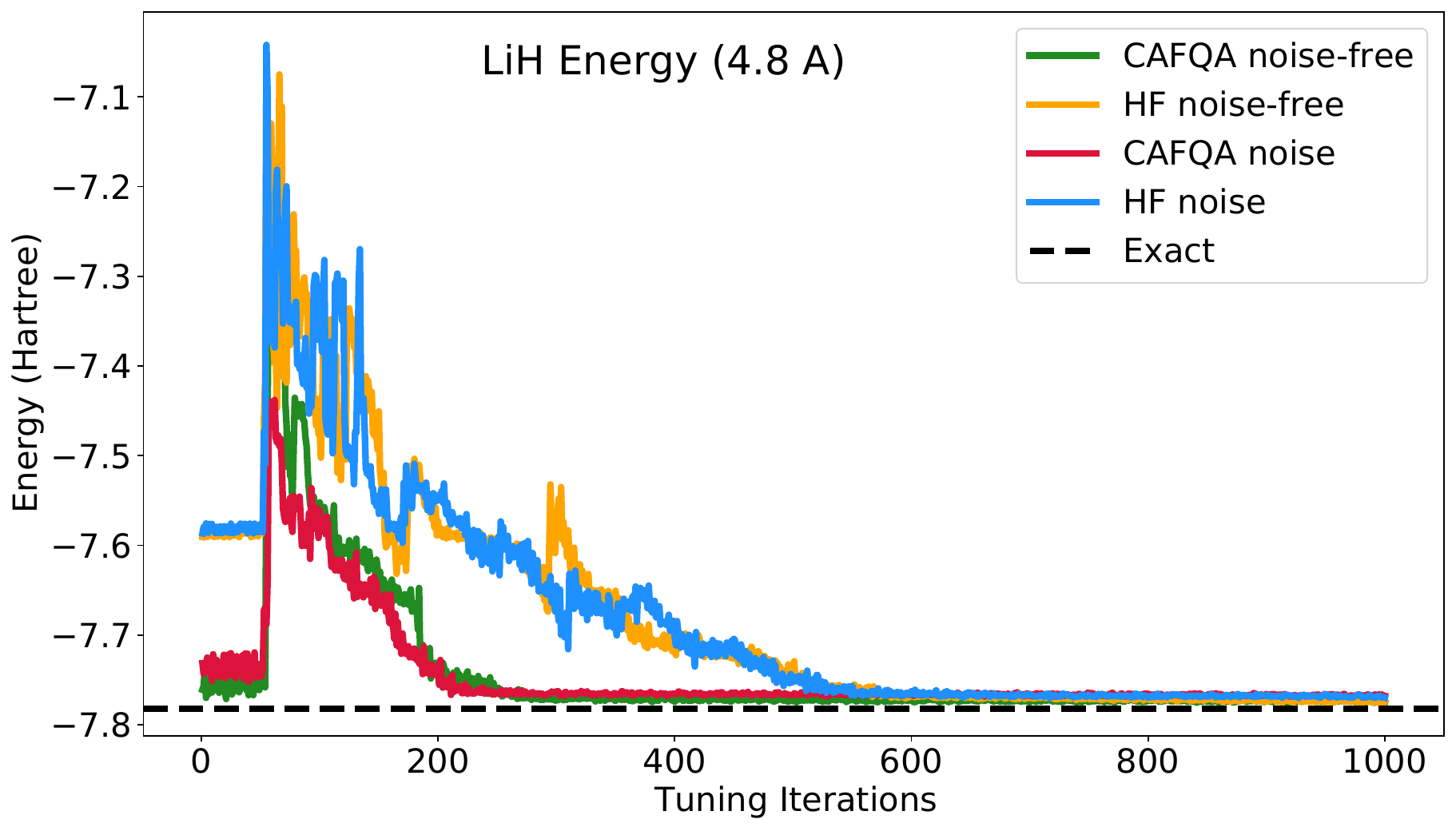}}
 
\caption{The plot shows the (a)  dissociation energy curve for LiH (Imported from  \cite{ravi2022cafqa} Fig. 9 (a))  and (b)  post-VQA tuning starting with different initialization (Imported from  \cite{ravi2022cafqa} Fig. 14.).
}
\label{Energy.LiH}
\label{fig:eval2}
\label{fig:post-cafqa}
\end{figure}

\noindent{\bf \emph{Numerical Analysis:}}
In our previous study \cite{ravi2022cafqa}, we conducted numerical simulations to estimate the ground state energy of molecules including H$_2$, LiH, H$_2$O, and H$_6$. For instance, in the case of LiH, \cref{Energy.LiH} (a) illustrates that CAFQA initialization produces an initial state that progressively recovers more correlation energy as LiH dissociates.
Continuing the VQE optimization process with CAFQA and HF initialization, we observed that CAFQA initialization leads to faster convergence compared to HF initialization in both noiseless and noisy simulations, as depicted in \cref{fig:post-cafqa} (b). A similar trend was observed for other molecular species tested in our study. However, this work was limited to simulation.

\subsection{Trapped-ion quantum computer}
\label{sec:background-TI}

Trapped-ion quantum computers are widely recognized among qubit platforms such as superconducting, silicon photonic, and neutral atoms. Several demonstrations of VQE have been successfully performed on trapped-ion quantum computers \cite{zhao2023enhancing, khan2023chemically, hempel2018quantum, nam2020ground, shen2017quantum}. One of the primary advantages of trapped-ion systems is their all-to-all connectivity, which allows any pair of ions within the system to perform native two-qubit gates without additional overhead. Many ansatz circuit compilation and optimization techniques achieve optimal results assuming this all-to-all connectivity \cite{linke2017experimental, wang2023ever, wang2021resource, nam2020ground}.
In addition to all-to-all connectivity, another promising feature of trapped-ion quantum computers is their scalability using the Quantum Charge-Coupled Device (QCCD) architecture \cite{kielpinski2002architecture}. QCCD enables the connection of multiple ion chips, effectively increasing the number of operational ions. More ambitious scaling schemes involve the use of photonic links to connect QCCD units \cite{brown2016co}.

In our study, we conducted VQE optimization on a trapped-ion quantum computer hosted by the Duke Quantum Center, Error-corrected Universal Reconfigurable Ion-trap Quantum Archetype (EURIQA) \cite{egan2021fault}. We leveraged the all-to-all connectivity during gate transpilation and implemented a hardware-efficient ansatz using minimal resources. For detailed technical insights into gate implementation and characterization outcomes, refer to \cref{sec:trapped-ion}.

\subsection{Hardware-software circuit interface}
\label{sec:bg-api}

From a top-level algorithm perspective, a quantum computer functions as a black-box that takes in a circuit and outputs a probability histogram. However, in practical applications such as running VQE, utilizing a quantum computer involves more than obtaining probabilities for a single circuit.
In VQE, the heuristic involves several steps: generating multiple circuits using a classical optimizer, executing these circuits on the quantum computer, performing post-processing on classical computing resources, feeding back expectation values to the classical optimizer, and iterating this process to refine results.

To facilitate this complex workflow, application programming interfaces (APIs) play a crucial role by seamlessly managing high-level circuit submission requests and translating them into low-level hardware-compatible commands.
Commercial software-hardware interfaces like Qiskit Runtime \cite{qiskitRuntime} exemplify this approach, enabling users to submit circuits with parameters along with necessary inputs such as Hamiltonian details and shot counts. These APIs are well-packaged and continuously maintained by dedicated teams.
In academic settings, pipelines are often tailored to specific environments, which may limit their applicability in more general scenarios and pose challenges in terms of code maintenance. For instance, at Duke Quantum Center, an open-source full-stack pipeline named QisDAX has been developed for trapped-ion devices \cite{riesebos2022modular,badrike2023qisdax}, integrating Qiskit with Duke ARTIQ (Advanced Real-Time Infrastructure for Quantum physics) extension \cite{bourdeauducq_2016_51303}. However, our specific trapped-ion quantum device could not directly utilize QisDAX due to compatibility issues with ARTIQ versions and required packages.

To address these challenges, we proposed a workflow that decouples high-level and low-level software dependencies. This approach simplifies the pipeline and makes it platform-independent, enabling academic researchers to quickly develop their own APIs for testing complex algorithms like VQE on diverse quantum hardware setups. For specific challenges encountered and detailed API workflow, refer to 
\cref{sec:api}.

\section{VQE enabled via the circuit interface}
\label{sec:api}
\begin{figure}
    \centering
\includegraphics[width=\linewidth]{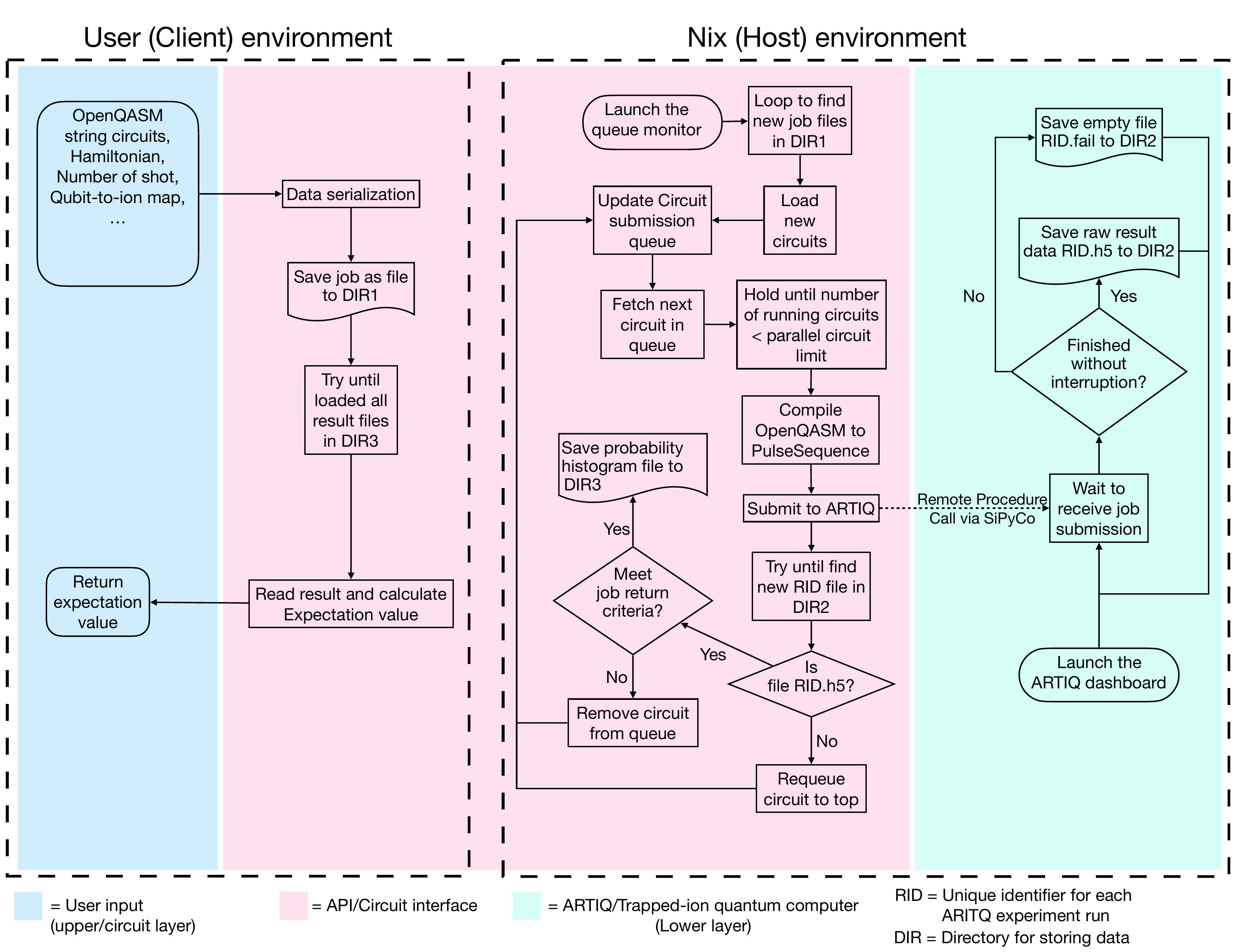}
    \caption{The workflow of the circuit interface API is designed to bridge the gap between the stable legacy Nix environment of the trapped-ion quantum computer and the diverse coding environments used by users. The API consists of two parts: one part is imported into the user's environment, and the other part is launched independently in the Nix environment and runs in the background. This setup allows the trapped-ion quantum computer to continue its development under a stable environment while providing users with the flexibility to work in any coding environment. The data exchange between the hardware layer and API layer is facilitated using `.h5' files, a hierarchical data format (HDF) used by ARTIQ to store raw data. 
 }
    \label{fig:api-workflow}
\end{figure}

\subsection{Workflow challenge in running VQE}

We have encountered many practical challenges on our way to demonstrate the CAFQA.
Here's a breakdown and discussion of these challenges:

\begin{enumerate}
    \item {\bf \emph{Incompatible software environment:}} 
    The compatibility between the circuit layer and hardware layer software environments poses a significant challenge. 
    For instance, the EURIQA trapped-ion quantum computer relies on ARTIQ-5 for integrating low-level hardware devices. To support this setup, a dedicated Nix environment was created, emphasizing robust hardware development conditions \cite{drewThesis2022}. As a result, software dependencies are rarely updated, such as the continuous use of legacy version Qiskit Terra (version 0.16.1). 
    However, this environment has limited compatibility with more versatile high-level circuit layer ecosystems. 
    This mismatch can complicate the integration and development process.

    \item {\bf \emph{API-hardware co-design:}} Efficiency in circuit execution is crucial and can be influenced by both software and hardware factors. ARTIQ, on the hardware end, manages job execution to minimize idling time. Optimizing the API to effectively utilize this feature is essential. Ensuring synchronization and efficient job management between the API and hardware can enhance overall performance.

    \item  {\bf \emph{Job return criteria:}}
    Handling unexpected hardware failures is a critical concern, especially during long VQE runs that can span days. Hardware errors may not be immediately detected, potentially leading to invalid data. Addressing these issues through robust job-return criteria is vital, particularly in academic settings where hardware is in continuous development and occasional interruptions are anticipated. Failing to manage these interruptions promptly can disrupt ongoing VQE iterations.
        
     \item  {\bf \emph{Per-circuit job queuing system:}} 
     Classical optimizers often generate multiple circuits for each iteration step of VQE, such as batches with the same ansatz but different measurement bases. When an unexpected hardware interruption occurs, it is ideal to rerun only the affected circuit rather than the entire batch generated for that step. 
     When the entire batch of circuits is regenerated by the classical optimizer, the entire batch of circuits must be compiled into pulse sequences and onto the FPGA one more time, significantly increasing the compilation overhead.
     As a result, implementing a per-circuit job queuing system helps minimize disruption and optimize resource utilization.

\end{enumerate}

These challenges are common across various quantum computing platforms, especially in academic laboratories with limited software expertise undergoing rapid development phases aimed at implementing APIs for running VQE-type circuits. Managing these complexities requires careful coordination between software and hardware layers, robust error handling mechanisms, and efficient job management strategies.

\subsection{Circuit Interface for VQE}
 Our goal is to design a pipeline that minimizes impact on the existing hardware development environment (host) and the high-level circuit layer (client). 
 To better facilitate running VQE, we propose a customized API for the EURIQA system trapped-ion machine, as illustrated in \cref{fig:api-workflow}.
Below we describe key design principles:

\begin{enumerate}
    \item {\bf \emph{Compatible data representation:}}
    Circuits and other metadata are transferred from client to host using the most compatible format. 
    This is essential when designing an API that bridges two distinct coding environments. 
    In our case, we utilize the OpenQASM string format \cite{cross2017open} to convey circuit information, as it can be recognized by both the legacy Qiskit Terra version 0.16.1 used in the host and more recent versions of Qiskit, Cirq, or any other quantum programming languages used by clients. Additionally, data is stored in a human-readable format for ease of transfer and troubleshooting.

    \item 
    {\bf \emph{Dual-end mechanism:}} 
    To bridge different code environments, the API consists of a client end and a host end. This approach effectively decouples the upper-level algorithm from occasional interruptions at the low-level hardware. As depicted in \cref{fig:api-workflow}, the client end of the API is imported as a regular module in the user's environment. It collects information from the classical optimizer, serializes the data, saves it to directory DIR1, and waits to load results from directory DIR3 saved by the host end of the API.
    Conversely, the host end of the API runs separately and independently under the lower-level software environment, continuously monitoring DIR1 in the background. When job return criteria fail multiple times, the host-end API halts and alerts the user to check the hardware. Once the hardware issue is resolved, the circuit queue resumes, starting with the failed circuit. Throughout the hardware failure event, the client-end API remains unaffected except for an extended waiting time to load all result files.
    To ensure the pipeline is straightforward and ease to debug, all circuit information passed to the host-end client is stored as files on disk. We use YAML (Yet Another Markup Language) format to store serialized data in a human-readable form. Transferring data via files offers the additional benefit of a simple network protocol between different environments. Since communication between the two ends of the API is file-based, a shared storage system, such as network-attached storage (NAS), is required. Laboratories commonly use NAS, but one can also use the file system of the same PC or connect the client-end and host-end APIs through other applicable network methods.

    \item 
    {\bf \emph{Highly customizable and future-proof:}} 
     The API is designed as a framework that can easily integrate with other tools. 
     For instance, the client-end API can incorporate the TriQ Toolflow \cite{murali2019full} for qubit mapping, gate optimization, and other tasks, updating the metadata before it is read. 
     Another example is integrating the API with device-level ARTIQ compilation optimization. 
     According to a recent article \cite{dalvi2023one}, a device-level partial-compilation (DLPC) can be applied at the ARTIQ-6 end to reduce repeated pulse compilation for parameterized VQE-type circuits.  The DLPC technique reduces classical compilation to be nearly constant or a linear factor of the memory available on the FPGA.
     To combine these methods, the API only needs to update the metadata accordingly and offload the "Compile OpenQASM to PulseSequence" step to the DLPC toolchain. 
    Additionally, due to the dual-end mechanism and text-based data representation, there is no restriction on the client-end coding language, making it more versatile. OpenQASM strings can also be replaced by other intermediate representations compatible with the host end.
    
\end{enumerate}

In addition to the main features above, there are some extra features tailored to our specific working environment that are worth explaining:

\begin{enumerate}
    \item 
    {\bf \emph{Parallel circuit submission:}}
    The API can control the number of jobs submitted to the ARTIQ dashboard. Typically, the client-end classical optimizer generates multiple VQE circuits to run, which are then queued at the host-end. The host-end can choose to submit all the circuits at once, one-by-one in sequence after receiving the experiment result of the previous circuit, or somewhere in the middle.
    When running a job actively with lasers, ARTIQ can prepare the next submitted job internally without accessing the hardware. In our case, without any modification to ARTIQ, keeping at least two jobs submitted to the ARTIQ dashboard instead of submitting jobs in sequence significantly improves overall circuit execution time, sometimes by as much as 60\%. However, we also do not want to overflow the ARTIQ dashboard by submitting all circuits in the API queue at once. As a result, we determined that the API should always maintain three jobs in the ARTIQ job stack at any time.

    \item 
    {\bf \emph{Job return criteria:}}
    The API supports a customized job return criteria. From test VQE runs, we observed patterns of several hardware failure events and concluded that such rare but significant failures can be conveniently characterized by an absurdly large deviation of the observed probability histogram from the target probability histogram. Since our circuit runs always use a small number of qubits, comparing with the exact probability histogram is trivial. Consequently, the host-end API examines the deviation and reruns the jobs before returning them to the client.
    \item 
    {\bf \emph{Job status monitor:}}
We have developed a graphical status monitor that visually represents real-time job submission statuses for the host-end API. This monitor displays key job states such as queuing, execution, and completion, offering users a clear and immediate view of experiment.

\end{enumerate}

This workflow can be easily modified to accommodate other platforms not using ARTIQ. To configure the API for a different backend, the major updates would involve replacing the ``Compile OpenQASM to PulseSequence'' and ``submit to ARTIQ'' steps with methods that match the desired backend. Additionally,  the corresponding post-processing method and job return criteria also need to be updated.
For example, besides the ARTIQ backend, we also built our own noisy simulator backend to test the job submission workflow and post-processing methods before running actual experiments.

\section{Trapped-ion quantum computer}
\label{sec:trapped-ion}

\begin{figure}
    \centering
\includegraphics[width=\linewidth]{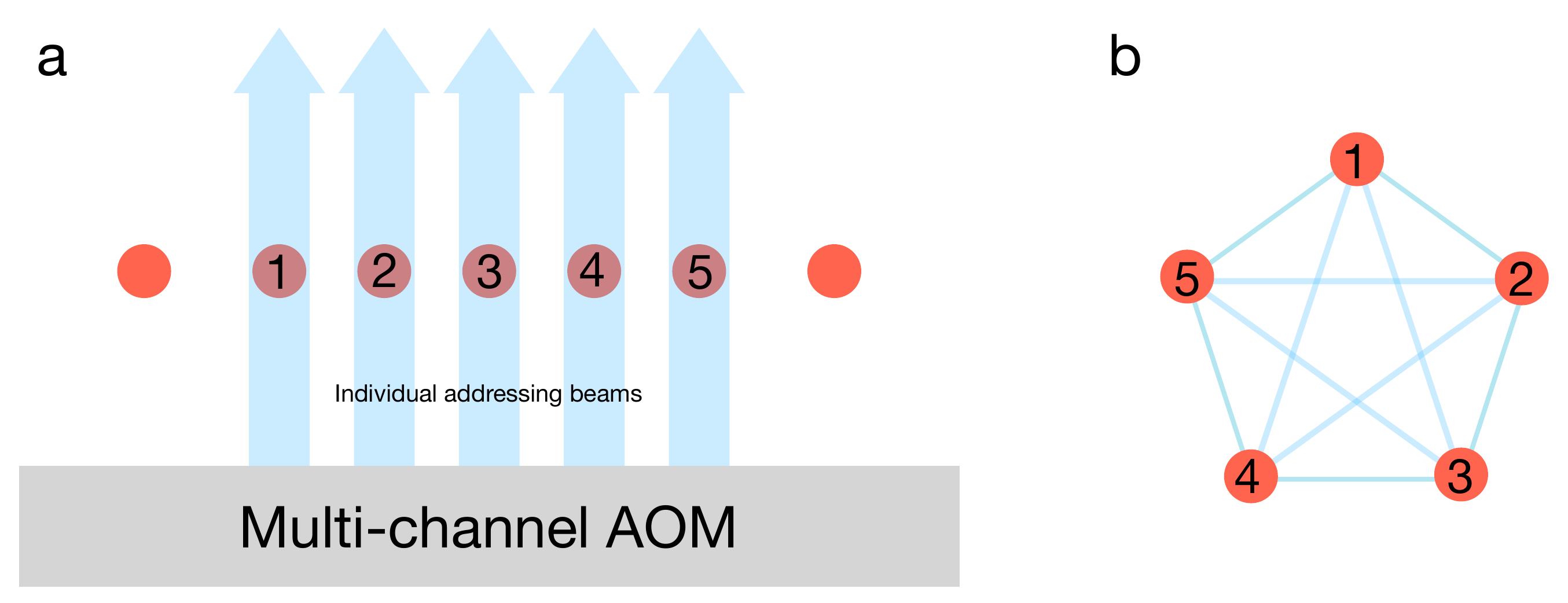}
    \caption{(a) The plot depicts a 7-ion chain, with the middle five ions identified as individually addressed working ions.
    (b) The plot illustrates all-to-all connectivity where each line represents an available $e^{-i\theta X_jX_k}$ two-qubit gate between vertices indexed $j$ and $k$. }
     \label{fig:trapped-ion}
\end{figure}

\noindent {\bf \emph{State preparation and readout:}}
The qubits in our trapped-ion setup are defined using the hyperfine levels of the ground state ${}^{2}S_{1/2}$ of $^{171}$Yb$^+$,
specifically $|0\rangle \equiv |F=0,m_F=0\rangle$ and $ |1\rangle \equiv |F=1,m_F=0\rangle$, which are separated by a frequency of 12.6 GHz. 
Qubit state measurement is achieved via state-dependent fluorescence, realized by the dipole-allowed ${}^{2}S_{1/2} \rightarrow {}^{2}P_{1/2}$ cycling transition \cite{noek2013high}.
This method allows reliable state readout by measuring the emitted fluorescence when the ion is excited.
Our state preparation and measurement (SPAM) error is 0.27(4)\%.

\noindent{\bf \emph{Individual ion addressing:}}
The trapped-ion quantum computer in this work can accommodate up to 32 addressable ions on a single chip, each representing a hyperfine qubit \cite{olmschenk2007manipulation}. 
The capacity is limited by the number of Acousto-Optic Modulator (AOM) channels in the control system. 
Previously, high-fidelity two-qubit have been demonstrated on a 15-ion chain using the same device \cite{egan2021fault}.
However, our current focus is on demonstrating the benefit of the CAFQA initialization method, which can be realized with fewer number of qubits. 
Longer ion chains demand more extensive and frequent calibrations to maintain performance, resulting in longer circuit execution times.
For our experiments, we use a configuration of seven ions arranged in a linear chain with individual laser beam addressing. In this setup, the five central ions are evenly spaced and serve as physical qubits. See \cref{fig:trapped-ion} (a) for the ion alignment.

\noindent{\bf \emph{Native gate operation:}}
Single-qubit gates are implemented using the composite SK1 pulse sequence \cite{PhysRevA.70.052318}, achieving a gate fidelity of 99.7(2)\% when applied to the central five qubits within the chain.
Two-qubit gates are realized using the M{\o}lmer-S{\o}rensen (MS) protocol \cite{sorensen1999quantum}, which entangles the internal states of ions with the collective vibrational modes of the ion chain. This is achieved through precise addressing of the relevant motional modes, leading to an effective native gate described by $XX(\theta) = \exp [-i\theta X_jX_k/2]$  for ion pair $j$ and $k$, where $X$ is the Pauli $X$ gate.
Unlike the CNOT gate, the 
$XX$ gate features a variable angle, making it particularly useful in certain VQE ansatz, such as the small-angle $XX$ gate application demonstrated for a water molecule \cite{ChemDemon4}. The all-to-all connectivity between any ion pair is realized by the shared phonon mode, as illustrated in \cref{fig:trapped-ion} (b).
To ensure robustness against frequency or amplitude noise of the driven pulses, amplitude modulation is performed via pulse shaping, detailed in ~\cite{debnath2016demonstration, egan2021fault}. Overall, the fidelity of our two-qubit gates varies between 99.3(1)\% and 98.9(2)\% (not SPAM corrected).

\section{Experiment demonstration}
\label{sec:CAFQA-exp}

\label{section:cafqa-demo}

\begin{table}
  \caption{Detail of the chosen molecular system and VQE setup. 
  Number of qubits is computed after frozen-core approximation and Qiskit's default qubit tapering technique with parity transformation. The number of measurement basis is obtained after qubit-wise commuting of the Hamiltonian.  }
  \label{tab:implementation detail}
  \begin{tabular}{lcc}
    \toprule
    Property &LiH &  BeH$_2$\\
    \midrule
    Point group &$C_{\infty v}$ &  $D_{\infty h}$ \\
    Bond length (\AA) &3.8   &2.8 \\
    Basis set           & \multicolumn{2}{c}{STO-3G}    \\
    Number of qubits&  \multicolumn{2}{c}{4} \\
    Number of measurement basis & 25  & 13 \\
    Shot number per circuit &\multicolumn{2}{c}{300} \\
    Budget for CAFQA BO & 1000   &3000\\
    VQE classical optimizer & \multicolumn{2}{c}{SPSA} \\
    SPSA learning rate budget &\multicolumn{2}{c}{25}  \\
    SPSA run budget & 400 &  350 \\
  \bottomrule
\end{tabular}
\end{table}

\begin{figure}
    \centering
\includegraphics[width=0.6\linewidth]{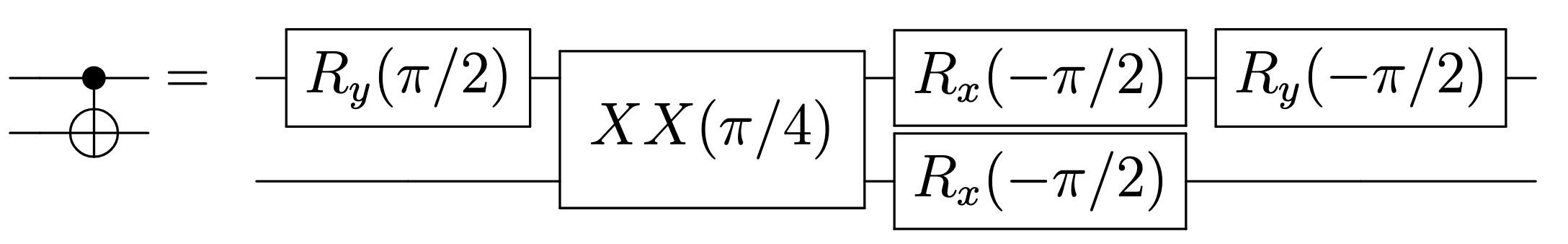}
    \caption{The CNOT gate can be converted to fixed-angle XX gate with necessary single-qubit gates. Later, these additional single-qubit gates can be absorbed into the parametrized single qubit gates in \cref{fig:hardware-efficient} (b). }
     \label{fig:cnot-conversion}
\end{figure}

\begin{figure}
    \centering
\includegraphics[width=0.8\linewidth]{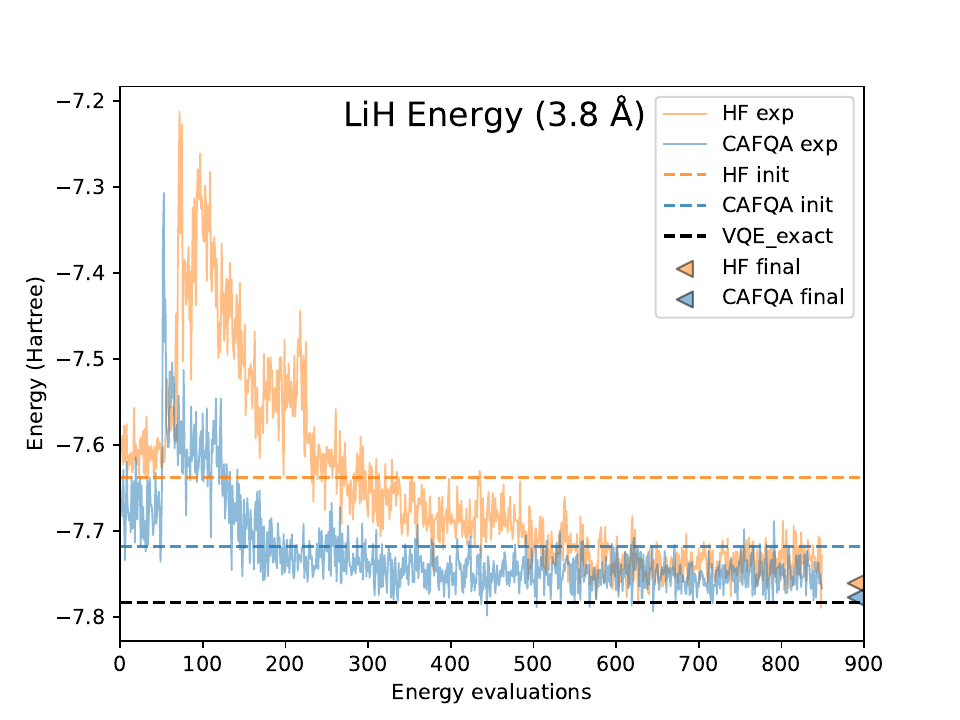}
    \caption{The plot illustrates the VQE optimization process of LiH energy starting from CAFQA (blue line) or HF (orange line) initialization on a trapped-ion quantum computer. 
    `CAFQA init' and `HF init' denote the initial energies from CAFQA and HF, respectively. 
    `VQE\_exact' represents the optimal VQE energy computed using a noiseless simulator with a sufficient number of iterations. For details on computing the final energies of CAFQA and HF, please refer to the main content. }
     \label{fig:LiH}
\end{figure}

\begin{figure}
    \centering
\includegraphics[width=\linewidth]{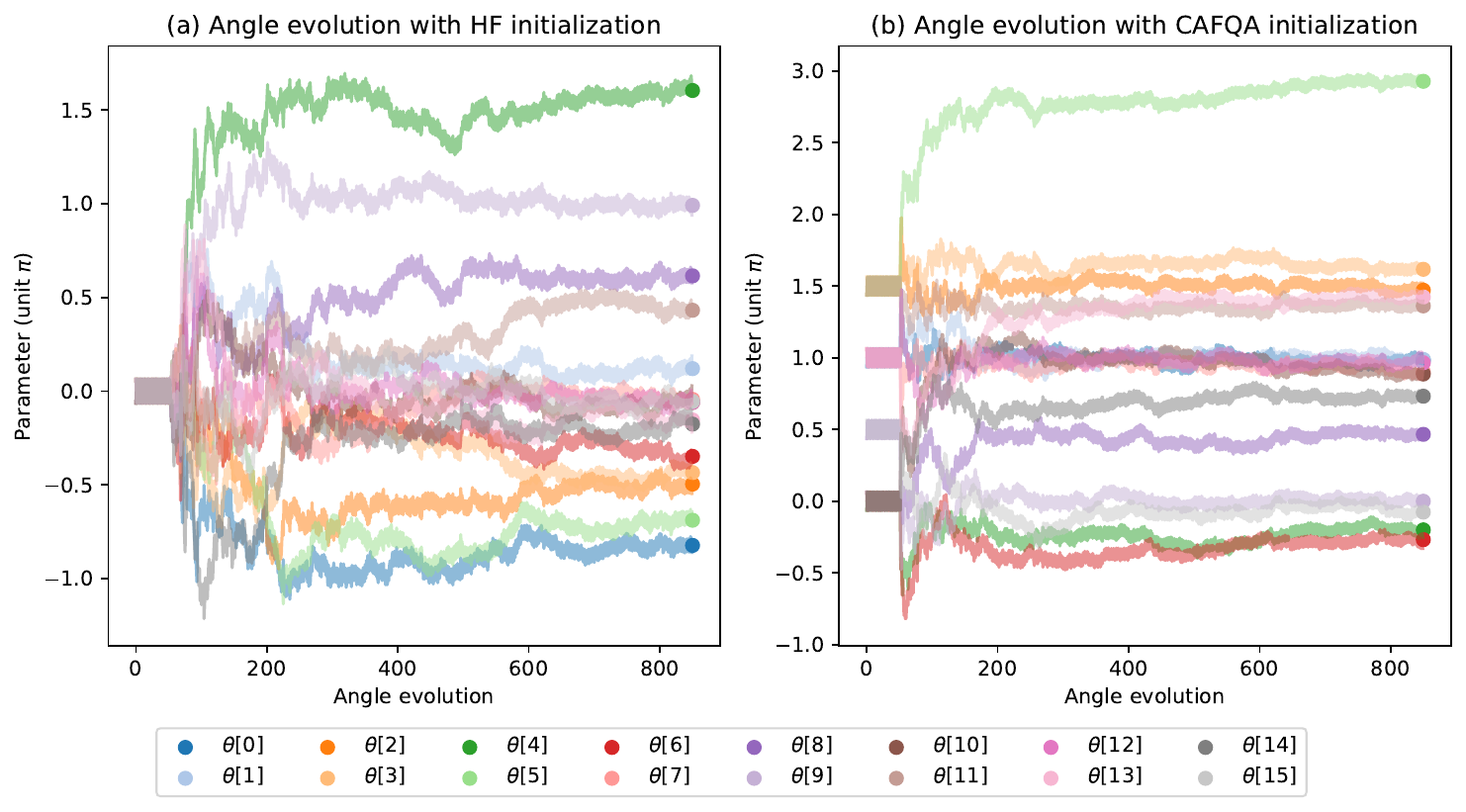}
    \caption{The plot displays the angle updates for LiH. 
    (a) Parameters starting with all 0 values represent the VQE with Hartree-Fock initialization. 
    (b) Parameters starting with different multiples of $\pi/2$ represent the VQE with CAFQA initialization. 
    The dots with the same color at the end of each initialization indicate the parameters averaged over the last 60 points. }
     \label{fig:LiH-angle}
\end{figure}

\begin{figure}
    \centering
    \includegraphics[width=0.8\linewidth]{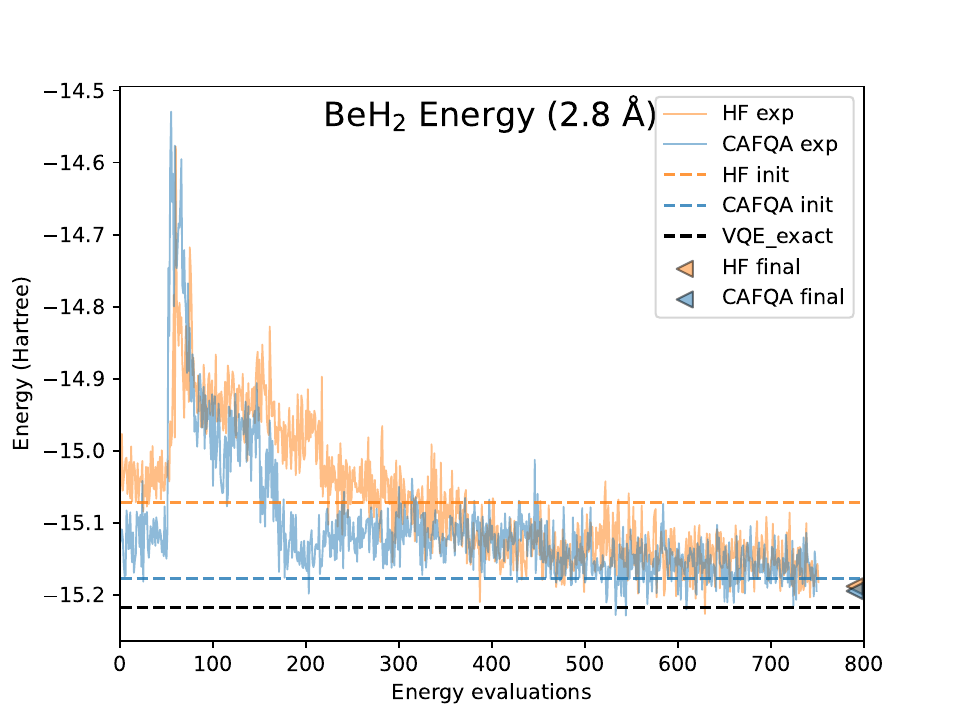}
    \caption{
    The plot illustrates the VQE optimization process of BeH$_2$ energy starting from CAFQA (blue line) or HF (orange line) initialization on a trapped-ion quantum computer. 
    `CAFQA init' and `HF init' denote the initial energies from CAFQA and HF, respectively. 
    `VQE\_exact' represents the optimal VQE energy computed using a noiseless simulator with a sufficient number of iterations. For details on computing the final energies of CAFQA and HF, please refer to the main content.}
     \label{fig:BeH2}
\end{figure}

\begin{figure}
    \centering
\includegraphics[width=\linewidth]{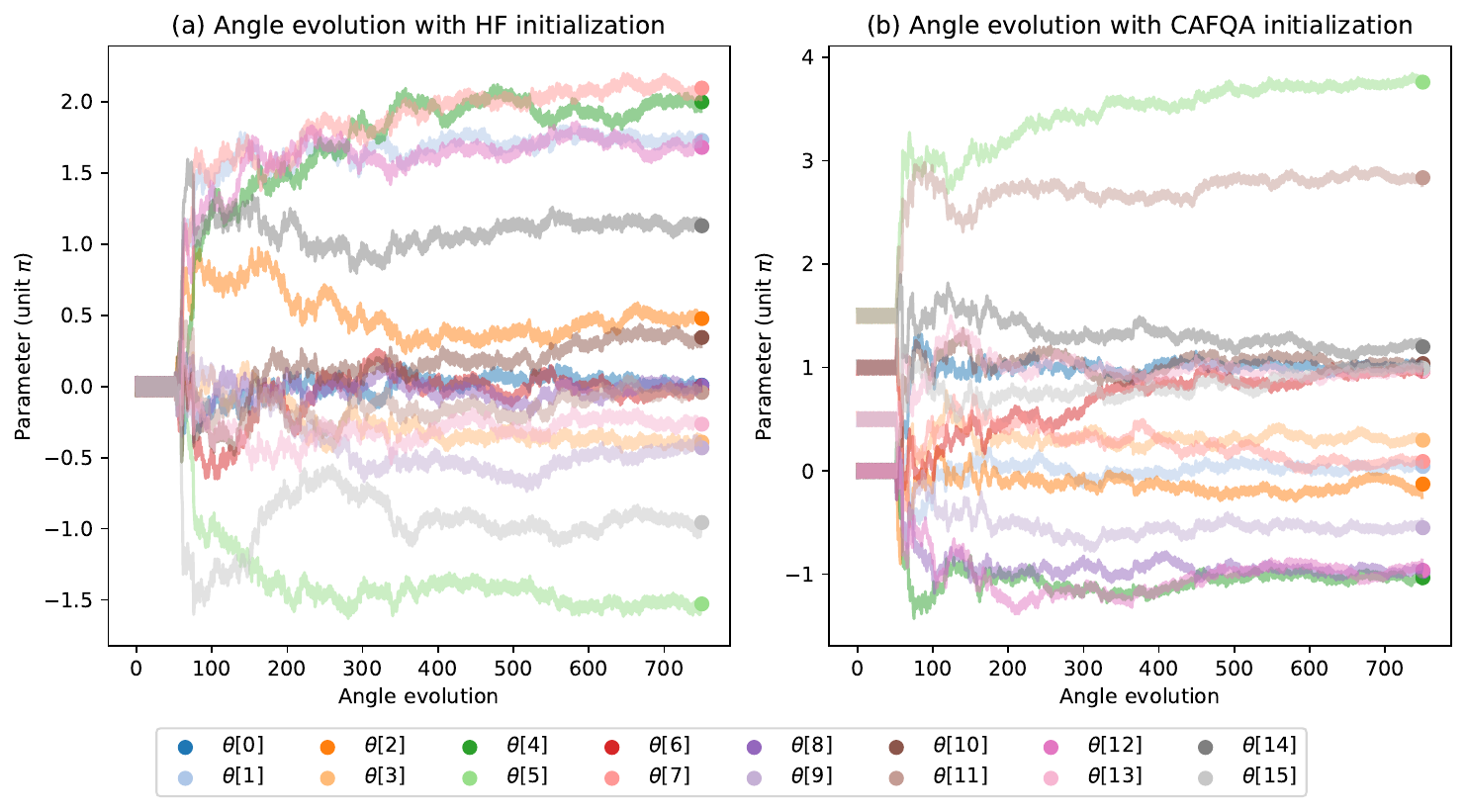}
    \caption{
    The plot displays the angle updates for BeH$_2$. 
    (a) Parameters starting with all 0 values represent the VQE with Hartree-Fock initialization. 
    (b) Parameters starting with different multiples of $\pi/2$ represent the VQE with CAFQA initialization. 
    The dots with the same color at the end of each initialization indicate the parameters averaged over the last 60 points. }
     \label{fig:BeH2-angle}
\end{figure}

We chose to demonstrate CAFQA with  LiH and BeH$_2$.
The specifications of the molecules and the VQE settings are listed in \cref{tab:implementation detail}.
To implement the VQE ansatz shown in \cref{fig:hardware-efficient} on our trapped-ion quantum computer, we convert the CNOTs into native $XX$ gates using the conversion rule in \cref{fig:cnot-conversion} and merged single-qubit rotations where applicable.
For VQE's classical optimizer, we employ the Simultaneous Perturbation Stochastic Approximation (SPSA) method, which is a gradient-based optimization technique widely used in VQE studies \cite{kandala2017hardware, spall1998overview, tran2024variational}. The SPSA method computes gradients by evaluating two energies at two perturbed parameter configurations, $E(\vec{\theta}^+)$ and $E(\vec{\theta}^-)$, where $\vec{\theta}^+$ and $\vec{\theta}^-$  are arrays of VQE parameters perturbed in the positive and negative directions, respectively.
For detailed calculations and methodology, refer to \cite{kandala2017hardware}.

The results for LiH are depicted in \cref{fig:LiH}. The horizontal axis represents the energy values obtained from each VQE energy evaluation. Notably, each VQE iteration involves two energy evaluations. The initial 50 evaluations consist of alternately computing $E(\vec{\theta}^+)$ and $E(\vec{\theta}^-)$, which are energies perturbed randomly around the starting point to estimate the learning rate. The computed gradient, along with the learning rate, facilitates the parameter updates for subsequent iterations. The SPSA algorithm then continues to iterate from the 51st evaluation onward.
The VQE energy plot in \cref{fig:LiH} conveys two key findings:
\begin{enumerate}
    \item   CAFQA initialization converges faster than HF initialization;
    \item  CAFQA initialization results in a lower final energy, indicating it recovers more correlation energy.
\end{enumerate}

The parameter optimization progress is illustrated in \cref{fig:LiH-angle}.
The plot shows that the parameters initialized with CAFQA  (\cref{fig:LiH-angle} (b))   stabilize  faster compared to those initialized with HF  (\cref{fig:LiH-angle} (a)). 
The SPSA algorithm eventually terminates at the budget limit.
The final parameters $\vec{\theta}_{\mathrm{final}}$ are computed by averaging the last 60 points, 
\begin{align}
\vec{\theta}_{\mathrm{final}} = \frac{1}{60}\sum_{j=s-29}^s (\vec{\theta}^+_j + \vec{\theta}^-_j),
\end{align}
where $s$ denotes the final step of the budget.
 These final parameters are indicated as dots at the end of the budget limit in  \cref{fig:LiH-angle}.
Subsequently, these parameters are used to compute the `CAFQA final' and `HF final' energies using a noiseless simulator, shown as triangles in \cref{fig:LiH}, to determine which initialization yields a lower final energy at the budget limit.

Similarly, the results for BeH$_2$ are presented in \cref{fig:BeH2} for energy convergence and \cref{fig:BeH2-angle} for parameter convergence. While the differences between HF and CAFQA are less significant in this example, the conclusions drawn from the LiH experiment are still applicable.

\section{Discussion}

\label{sec:discussion}

\noindent{\bf \emph{Plan for more extensive study:}}
Although we have successfully demonstrated the faster convergence of CAFQA initialization compared to HF initialization within the given budget, more extensive studies are necessary to thoroughly evaluate the capability of CAFQA on real quantum devices.
First, a wider variety of molecules should be sampled, especially larger systems. For the size of problems that we are able to demonstrate in this work, there is less opportunity to be gained from novel initialization. If our results from simulations translate to real devices, we expect more significant CAFQA benefits from larger molecular systems.
It should also be noted that benefits of initialization become less significant if devices are too noisy; as quality of academic devices improve, we expect to see more favorable convergence trends. 
For each molecule, there are also many degrees of freedom in bond length and the basis set. These configurations will sample different types of Hamiltonians. Moreover, within these configurations, there typically exists a large number of degenerate CAFQA initializations. However, this degeneracy might break during subsequent VQE continuous-variable evolution. Predicting which CAFQA initialization will result in better convergence remains an open topic for future research. To understand how degenerate CAFQA initializations break degeneracy, it is necessary to repeat many complete VQE optimizations for each degenerate CAFQA initialization. This is due to the stochastic nature of SPSA and the noise in quantum devices, which requires comparing the spread of the VQE convergence curves. 
However, the systematic tests proposed above are extremely time-consuming and beyond the capability and schedule availability of the current academic device. The primary goal of this article, in terms of CAFQA demonstration, is to show that the advantage of CAFQA initialization is practical on real quantum devices, which is clearly evident in the data we have presented.

\noindent{\bf \emph{Run experiment with API:}} 
In total, the experiments LiH and BeH$_2$ needs to run 42,500 and  19,500 unique circuits, respectively. 
These circuits were executed intermittently over approximately two weeks, with many hardware interruptions and necessary maintenance in between. 
The API designed as shown in \cref{fig:api-workflow} proved to be robust and effective, automating the workflow to allow circuits to run overnight with minimal interference. The API has become an essential tool for executing gate-model circuits and is facilitating many other projects currently running on the EURIQA TI machine.

\section{Conclusion}
In this study, we have demonstrated the advantage of CAFQA initialization over the traditional Hartree-Fock initialization using a trapped-ion quantum computer. While we have only tested a few cases, the convergence plots for LiH and BeH$_2$ clearly show that VQE convergence can benefit from better initialization. These experiments can serve as a reference for more complex VQE experiments requiring improved initialization methods. We hope that the demonstration presented here will inspire future VQE experiments, allowing CAFQA to be tested on different hardware platforms and optimization problems.
The trapped-ion quantum computer used in this study is hosted in an academic setting. To facilitate the demonstration, we designed a workflow featuring a dual-end mechanism to pass necessary information between two independent coding environments. This workflow allows for running VQE experiments in an academic environment with high tolerance for hardware interruptions, providing a practical framework for executing more sophisticated circuits in the future.

\begin{acknowledgments}
This work is supported by the ARO through the IARPA LogiQ program;
the NSF QLCI program;
STAQ under award NSF PHY-1818914/232580;
the DOE QSA program;
the AFOSR MURIs on Dissipation Engineering in Open Quantum Systems, Quantum Measurement/ Verification, and Quantum Interactive Protocols;
the ARO MURI on Modular Quantum Circuits;
the DOE HEP QuantISED Program;
and in part by the US Department of Energy Office of Advanced Scientific Computing Research, Accelerated Research for Quantum Computing Program.
This research used resources of the National Energy Research Scientific Computing Center, a DOE Office of Science User Facility using NERSC award NERSC DDR-ERCAP0030278.
FTC is the Chief Scientist for Quantum Software at Infleqtion and an advisor to Quantum Circuits, Inc.
\end{acknowledgments}

\bibliographystyle{ACM-Reference-Format}
\bibliography{reference}

\end{document}